\documentclass[prb,twocolumn,showpacs,amsmath,amssymb]{revtex4}

\usepackage{graphicx}       
\usepackage{dcolumn}        

\newcommand{\beq}{\begin{equation}}
\newcommand{\eeq}{\end{equation}}
\newcommand{\beqa}{\begin{eqnarray}}
\newcommand{\eeqa}{\end{eqnarray}}

\newcommand{\kvec}{{\bf k}}

\newcommand{\qvec}{{\bf q}}

\newcommand{\OO}{{\overline \Omega}}

\begin{document}
\title{Unravelling the glue and the competing order in superconducting cuprates}
\author{S. Caprara$^1$, C. Di Castro$^1$, B. Muschler$^2$,
R. Hackl$^2$, M. Lambacher$^2$, A. Erb$^2$, S. Komiya$^3$, Y. Ando$^4$, and M. Grilli$^1$}

\affiliation{$^1$CNR-INFM-SMC, and Dipartimento di Fisica, Universit\`a di 
Roma ``La Sapienza'', P.$^{le}$ Aldo Moro 5, 00185 Roma, Italy}
\affiliation{$^2$Walther Meissner Institut, Bayerische Akademie der Wissenschaften, 
85748 Garching, Germany}
\affiliation{$^3$ {Central Research Institute of the Electric Power Industry, Komae, 
Tokyo 201-8511, Japan}}
\affiliation{$^4$ {Institute of Scientific and Industrial Research, Osaka University, 
Ibaraki, Osaka 567-0047, Japan}}

\begin{abstract}
{
We present Raman scattering 
experiments in ${\rm La_{2-x}Sr_xCuO_4}$ single crystals at various doping levels $x$ 
and compare the results with theoretical predictions obtained assuming an interaction
mediated by spin and charge fluctuations. The light-scattering selection rules allow 
us to disentangle their respective contributions. We find that the glue spectral 
function is spin-dominated at low doping while the contribution of charge fluctuations 
becomes dominant around optimal doping. This indicates that the fluctuations of 
a nearly ordered state with coexisting spin and charge order
support the superconducting pairing.}
\end{abstract}
\date{\today}
\maketitle

\section{Introduction}
Several systems, ranging from heavy fermions to manganites, from high-temperature
superconducting cuprates to ruthenates, display anomalies in the metallic phase, with the 
violation of the well established paradigm of Fermi-liquid theory. 
As a matter of fact all these systems are characterized by strong 
electron-electron correlations
and display similar phase diagrams, where an ordered phase is adjacent to the metallic
state, possibly with the occurrence of superconductivity.
This raises the question about the effective electron-electron interactions determining
the metallic anomalies and possibly (high-temperature) superconductivity.
In this respect cuprates are a paradigmatic example:
Around optimal doping (where the superconducting critical temperature $T_c$ 
is highest) these systems display 
anomalies in spectroscopic, transport, and thermodynamic properties. These anomalies 
should stem from the same effective interactions which also provide pairing.
Understanding their nature is the hotly debated ``glue issue''. For some people 
\cite{glue,anderson1,LNW} the glue consists of strong, essentially instantaneous, 
electron-electron interactions, arising from the doped Mott insulator character of 
these systems. According to a second point of view, the low-energy effective 
interaction is due to a retarded bosonic-like glue \cite{scalapino,hanke}. Besides 
the standard phonons, spin waves \cite{chubukov} are the most quoted candidates, 
as reminiscent of the antiferromagnetism suppressed by doping. Quite naturally, 
the bosonic excitations might also be related to an elusive 
electronic order, which competes with superconductivity and should occur in 
underdoped cuprates (i.e. at doping less than optimal) either 
as a long-range ordered phase or in the form of 
local/dynamical short-range fluctuations 
\cite{CDG,varma1,varma2,reviewQCP1,reviewQCP2,benfatto,chakravarty,metzner}.
The proximity to the corresponding ``critical region'' naturally brings along 
abundant critical fluctuations and leads to
strongly momentum-, temperature- and doping-dependent bosonic effective
interactions. Within this second point of view the identification of the glue also 
sheds light on the underlying competing instability.

Here we show that exploiting the specific properties of Raman spectra in the different symmetry 
channels, fluctuations with different characteristic wavevectors can be identified.
This is a quite general result for effective interactions that are
strongly peaked at a finite wavevector (see Appendix  for technical details) 
and it might render Raman spectroscopy a particularly powerful tool
to detect and study elusive orders occurring at finite wavevectors.
More specifically, here we 
start from a phenomenology for spin and charge fluctuations, predict the Raman 
spectra in the two $B_{1g}$ and $B_{2g}$ symmetry channels 
and compare our theoretical results with experimental results on ${\rm La_{2-x}
Sr_xCuO_4}$ (LSCO) samples at various doping levels $x$ and temperatures $T$.  In this way
 we disentangle the contribution of spin and charge fluctuations, which
can be identified as the relevant bosonic modes in these 
systems. This identification, not only clarifies the glue issue, but 
also strongly indicates a nearly spin/charge-ordered state as the competing phase 
in underdoped cuprates, which evolves from a dominating spin order at low doping, near 
the Mott insulator, to a charge-dominated order on the overdoped side.

\section{Experimental details}
The samples were grown using the traveling solvent floating zone (TSFZ) method. The characterization is 
presented in Table~\ref{tab:samples}. 
\begin{center}
\begin{table}
\caption[]{List of samples. Samples labeled with $a$ have been prepared by M. Lambacher and A. Erb (WMI 
Garching) \cite{Lambacher:2010}, $b$ by Seiki Komiya and Yoichi Ando (CRIEPI, Tokyo  and Osaka University) 
and $c$ by N. Kikugawa and T. Fujita (Hiroshima and Tokyo). The transition temperatures were measured either 
resistively or via magnetometry or via the non-linear ac response.}
\centering
\begin{tabular}{c c c c c c}
 \hline\noalign{\smallskip}
 sample  & doping $x$ & ~$T_c$ (K) & $\Delta T_c$ (K)&  comment & \\
 ${\rm La_{1.85}Sr_{0.15}CuO_{4}}$                    & 0.15 & 38    & 3 & O$_2$ annealed & $a$ \\
 ${\rm La_{1.83}Sr_{0.17}CuO_{4}}$                    & 0.17 & 39    & 1 & O$_2$ annealed & $b$ \\
 ${\rm La_{1.80}Sr_{0.20}CuO_{4}}$                    & 0.20 & 24    & 3 & O$_2$ annealed & $a$ \\
 ${\rm La_{1.75}Sr_{0.25}CuO_{4}}$                    & 0.25 & 12    & 3 & O$_2$ annealed & $a$ \\

  ${\rm La_{1.74}Sr_{0.26}CuO_{4}}$                   & 0.26 & 0     & - & O$_2$ annealed    & $c$ \\

\noalign{\smallskip}\hline
\end{tabular}
  \label{tab:samples}
\end{table}
\end{center}
The lowest doping was $x=0.15$ close to the $T_c$ maximum where 
Landau-Fermi liquid theory should still be applicable. The overdoped sample with $x=0.26$ had no 
indication of superconductivity above 2\,K. All samples were post-annealed in pure oxygen at 1\,bar 
to improve the crystal quality. The spectra were taken on polished surfaces. In the case of 
${\rm La_{1.74}Sr_{0.26}CuO_{4}}$ the results were compared to those from a cleaved surface and were found 
to be identical to within the experimental error \cite{Muschler:2010}.

The Raman experiments were performed with standard equipment. For excitation an Ar 
ion laser operated at 
458\,nm was used. The angle of incidence of the incoming photons was 66$^{\circ}$. 
The polarization state 
outside was prepared in a way that photons inside the sample had the desired state. 
Scattered light of a 
selected polarization was collected along the surface normal and focused on the 
entrance slit of the 
spectrometer. The energy-selected photons were detected with a liquid nitrogen 
cooled CCD detector. All 
spectra are corrected for the sensitivity of the complete setup. We generally show 
the imaginary part of 
response functions, $\chi_{\mu}^{\prime\prime}$, at pure symmetries $\mu=B_{1g}$, 
$B_{2g}$ which are related to the measured cross section as
\begin{equation}
\frac{d^2\sigma_{\mu}}{d\tilde{\Omega} d\omega_s} = A\frac{\omega_i}{\omega_s}\{1+n(\omega,T)\}
\chi_{\mu}^{\prime\prime}({\bf q}=0, \omega,T)
\end{equation}
with $\sigma_{\mu}$ the photon cross section in symmetry $\mu$, $\tilde{\Omega}$ 
the solid angle accepted by 
the collection optics $\omega=\omega_i-\omega_s$ the energy transferred to the system, $A$ a constant 
absorbing all factors to convert $\chi_{\mu}^{\prime\prime}$ into a cross section and 
$n(\omega,T)=[\exp(\omega/T)-1]^{-1}$ the 
Bose-Einstein factor.  To simplify the notation we drop
hereafter the index $\mu$, since the formal expressions in the following
 are the same for all symmetry channels.

{\section{Raman spectra}}

\begin{figure*}
\includegraphics[angle=0,scale=1.]{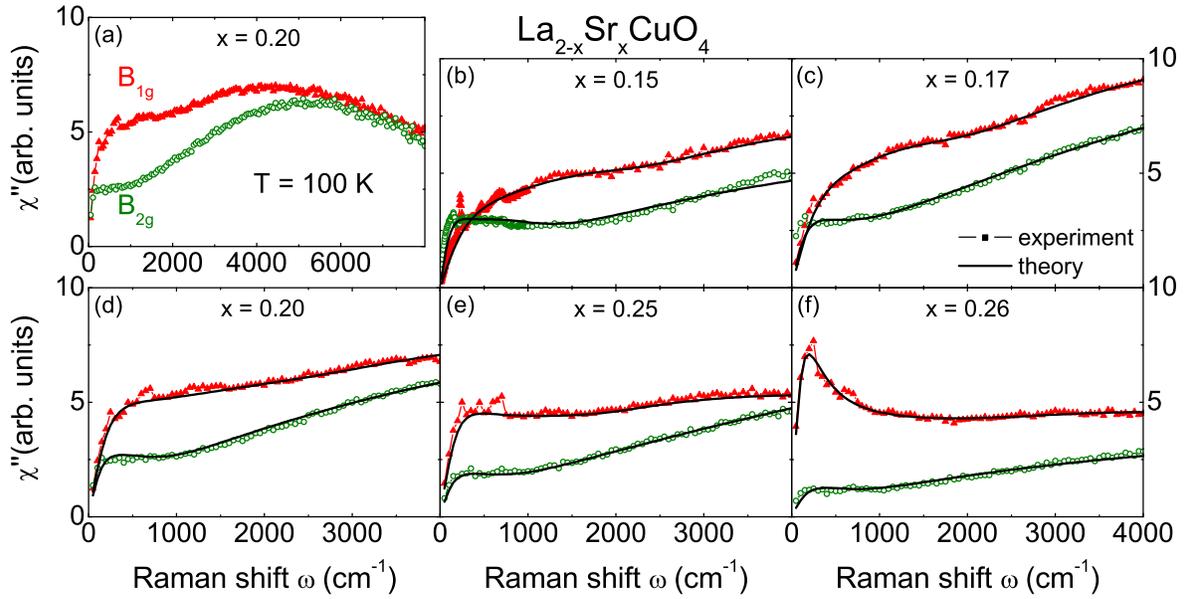}
\vspace{0.5 truecm}
\caption{(a) Experimental Raman response $\chi''_\mu(\omega,T)$ of LSCO at $x=0.20$ on 
a broad
frequency range up to 8000 cm$^{-1}$ at intermediate $T=104$\,K in the $B_{1g}$ channel 
(green circles)
and in the $B_{2g}$ channel (red triangles). (b-e)
Raman response $\chi''_\mu(\omega,T)$ of LSCO at $x=0.15,0.17,0.20,0.25$,  and $0.26$
on a reduced frequency range up to 4000 cm$^{-1}$.
Data in the $B_{1g}$ channel are given by green circles, while the
spectra in the  $B_{2g}$ symmetry are given by red triangles. The various spectra are taken 
at
similar temperatures around $T~100$K. The results of the fits obtained within the
``nearly-critical'' charge/spin theory (see text) are given by the black solid lines.
}
\label{fig1}
%
\end{figure*}

\begin{figure*}
\includegraphics[angle=0,scale=1.]{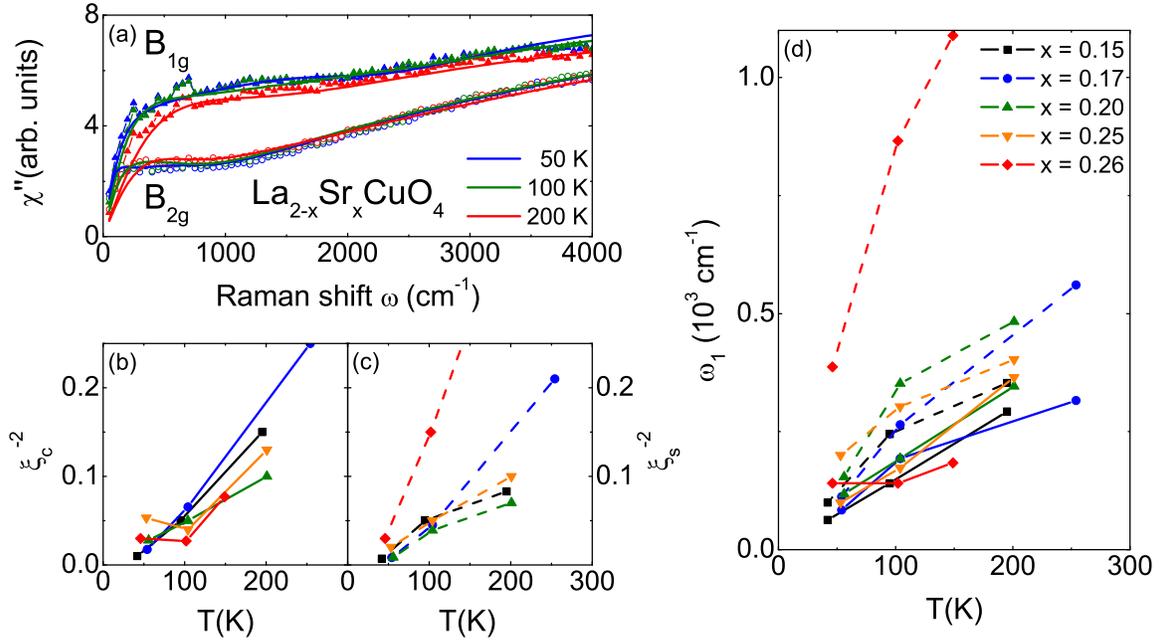}
\vspace{0.5 truecm}
\caption{(a) Raman response $\chi''_\mu(\omega,T)$ of LSCO at $x=0.20$ in the
$B_{1g}$ (circles) and $B_{2g}$ (triangles) at three
different temperatures above $T_c$ and the related fits (solid curves) 
obtained within the nearly critical
scheme described in the text: $T=56$\,K (blue circles and curve), $T=104$\,K 
(green circles and curve),
$T=201$\,K (red circles and curve). (b) Temperature dependence of the inverse square
correlation length $\xi^{-2}_c$ of the charge modes
at different dopings ($x=0.15$ blue, $x=0.17$ red, $x=20$ green, $x=0.25$ black). 
(c) Same as in
(b) for the inverse square correlation length
$\xi^{-2}_s$ of the spin collective modes. (d) Temperature dependence 
of the low-frequency typical energy scale
$\omega_1=\sqrt{m\overline{\Omega}}$ of the charge and spin collective modes.
}
\label{fig2}
%
\end{figure*}

\noindent
Electronic Raman scattering is a bulk (nearly surface-insensitive) probe and it
measures a response function $\chi^{\prime\prime}(\omega,T)$ analogous to that of the optical
conductivity \cite{Shastry:1990}.
However, while the latter yields an average over the entire Brillouin zone,
light scattering
projects different parts of the Brillouin zone for different polarizations
of the incoming and outgoing photons \cite{Devereaux:2007},
thereby introducing specific form factors in
$\chi^{\prime\prime}(\omega,T)$. We consider Raman
spectra in the $B_{1g}$  and $B_{2g}$ channels obtained from the systematic 
analysis of several LSCO samples
at different doping levels $x$ ranging from the nearly optimal up to the 
strongly overdoped region (see
Fig. \ref{fig1}). 
Generally we observe that the overall shape of the spectra in $B_{1g}$ and $B_{2g}$
channels is quite different up to 3000-4000 cm$^{-1}$,
while at higher frequency the two spectra have similar shape and intensity.
At frequencies up to 2000 cm$^{-1}$, the $B_{1g}$ spectra can be schematized by a hump
appearing as a shoulder of a slightly larger hump peaked at
 $\omega \sim 4000$ cm$^{-1}$.
On the other hand, $B_{2g}$ spectra display a plateau
up to about $1000$ cm$^{-1}$. At higher frequencies the absorption rises
and a hump nearly twice as large, similar to the one found in $B_{1g}$,
appears at about $4000$ cm$^{-1}$.
The temperature dependence of the spectra is altogether weak (see Fig. \ref{fig2}a),
but it may even be opposite in the two channels in an intermediate frequency range.
In particular the temperature dependence of the $B_{2g}$ spectra corresponds to those of the 
optical conductivity for $x > 0.05$. In $B_{1g}$ symmetry at low-intermediate dopings
there is a 
strong extra contribution from fluctuations at low frequency\cite{tassini,suppa,Muschler:2010}. 
Therefore we confine the analysis to $x \ge 0.15$.

A first remark is now in order: Although the Raman form factors select different regions of the
Brillouin zone, and therefore explore different electronic structures,
it is rather unlikely that the changes in doping
and temperature in the two channels
[see the low-energy part of Figs. 2(a) and the overall shapes of the responses in Figs. 1(b-e)]
can simply be attributed to diverse evolutions of the underlying fermionic band structures.
Also a scattering mechanisms
with zero characteristic momentum \cite{varma1,varma2,LNW}, would act in the same way in the
two channels hardly explaining their different behaviours.
Guided by a wealth of experimental evidences for a
prominent tendency of LSCO cuprates to form charge-ordered states
(stripes or checkerboard), we show below that
different shapes (and different temperature and doping evolutions)
of the spectra in the two channels naturally arise from quasiparticles
(QPs) coupled to distinct
charge and spin fluctuations, each one having
its own dynamics and characteristic wavevector.

\vspace{0.5 truecm}
\noindent
{\section{The ``glue'' collective modes}}

\noindent
Here we approach the identification of the glue extracting its properties
from the Raman data. We start from a form of damped glue boson collective modes (CMs), 
which is customary for
spin fluctuations \cite{mmp,chubukov}
\beq
D_\lambda({\bf q}, \omega)=-\frac{1}{m_\lambda+\nu_\lambda
({\bf q}-{\bf q}_{\lambda})^2-i\omega-
\omega^2/{\OO_\lambda}} \label{propagator}
\eeq
where $\lambda=c,s$ refers to charge or spin CMs
substantially peaked at characteristic wavevectors $\qvec_s\approx (\pi,\pi)$ and
$\qvec_c\approx (\pm \pi/2,0),(0,\pm \pi/2)$ respectively.
Here the mass $m_\lambda$ is the minimum energy required to excite the CM and
$\nu_\lambda$ is a fermion scale setting the CM momentum dispersion. This dispersion is
limited by an energy cutoff $\Lambda_\lambda$. The dimensionless quantities $m_\lambda/\Lambda_\lambda$ are
the inverse square correlation lengths (in units of the lattice spacing), which measure
the typical size of ordered domains.
The $i\omega$ term establishes the low-energy diffusive character of
these fluctuations due to decay into particle-hole pairs, whereas above
the scale set by $\OO_\lambda$ the CM has a more propagating character.

\begin{figure}
\includegraphics[angle=0,scale=0.6]{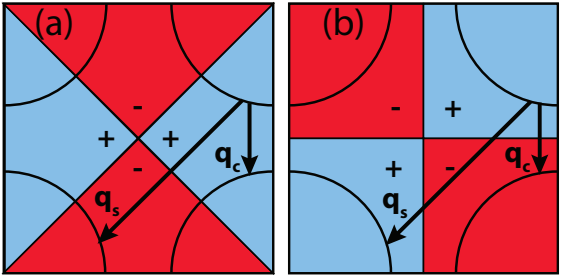}
\caption{Hot spot structure for both charge (with $\qvec_c$) and spin  (with $\qvec_s$)
scattering. In panels (a) and (b) the sign structures of the form factors for
the  $B_{1g}$ and $B_{2g}$ channels, $\gamma_{1g}(\kvec)=\cos(k_x)-\cos(k_y)$ and
$\gamma_{2g}(\kvec)=\sin(k_x)\sin(k_y)$
(we use a unit lattice spacing), respectively, are reported.
}
\label{cmsr}
\end{figure}

The CMs mediate strongly momentum-dependent scattering
between the QPs. At low energy this scattering mechanism is more effective
for QPs on the Fermi surface, which are connected by the characteristic CM momenta
(the so-called ``hot spots''). For the typical $\qvec_s$ and $\qvec_c$
in the cuprates the hot spots for spin and charge
CMs occur in the same regions of the Fermi surface near the $(\pm \pi,0)$ and
$(0,\pm \pi)$ points of the Brillouin zone (see Fig. \ref{cmsr}).

The Raman response function can be expressed in terms of a memory function 
as
\beq
\chi(\omega)=\frac{\chi_0\omega}{\omega+M(\omega)}.
\label{chi}
\eeq
Here $\chi_0$ is the purely real low-frequency Raman response function in the absence of any scattering process. 
As in standard approaches \cite{mahan,goetze}, our perturbative memory function calculation is limited to
processes which involve the exchange of one CM ($\lambda=c$ or $s$).
The specific form of the fluctuations (\ref{propagator}) allows one to
identify and analytically calculate the dominant contribution to $M=M_c+M_s$
(for details see the Appendix) yielding for the imaginary part
\beqa
&&ImM_\lambda(\omega)=\frac{1}{\omega} \int_0^\infty dz \left[ \alpha^2 F_\lambda(z)\right]
\left[2\omega \coth\left(
\frac{z}{2T}\right) \right. \nonumber \\
 &&-\left.(z+\omega)\coth\left(\frac{z+\omega}{2T}\right) +(z-\omega)
\coth\left(\frac{z-\omega}{2T}\right) \right]
\label{immemory}
\eeqa
which has the meaning of a frequency-dependent inverse scattering time.
The crucial physical ingredient
in Eq. (\ref{immemory}) is the spectral distribution $\alpha^2F$
appearing, e.g., in the textbook Eliashberg theory of superconductivity
(the so-called ``glue function''), which in our case reads
\beqa
\alpha^2F_\lambda(\omega)&=&g_\lambda\left[
\arctan\left(\frac{\Lambda_\lambda\OO_\lambda-\omega^2}{\OO_\lambda\omega} \right)\right.\nonumber \\
&-&
\left.\arctan\left(\frac{m_\lambda\OO_\lambda-\omega^2}{\OO_\lambda\omega} \right)\right]
\label{glue}
\eeqa
The dimensionless parameter $g_\lambda$, which also absorbs $\nu_\lambda$,
describes the coupling between CMs and QPs.
This spectrum is roughly contained in the range between
$\omega_1\approx \sqrt{\OO m}$ and $\omega_2\approx \sqrt{\OO \Lambda}$.
When $\OO$ is large, $\alpha^2F(\omega)$
has the broadly peaked shape of a glue function due to diffusive CMs.
For small $\OO$,
i.e., for rather propagating CMs, $\alpha^2F(\omega)$ has a ``flattish'' box-like shape.
This latter form of the scattering spectrum has been phenomenologically assumed since the early
days of high-$T_c$ superconductivity by the marginal-Fermi-liquid theory \cite{varmaMFL,abrahamsMFL}.
We find it non trivial that the marginal Fermi
liquid phenomenology in momentum-integrated quantities (tunnelling spectra, optics, Raman, etc.)
is obtained here from a microscopic strongly momentum dependent scattering mechanism.
When adapted to the calculation of optical conductivity, our formalism
contains as limiting cases the marginal-Fermi liquid and the diffusive CM
considered in Ref. \onlinecite{normanchubukov} and reproduces their results.

Our first general result is that the different experimental spectra in the $B_{1g}$ and
$B_{2g}$ channels are obtained and correspond to a more diffusive and a more propagating
CM respectively (an extended presentation of this generic result
is given in the Appendix).

\vspace{0.5 truecm}
\noindent
{\section{Analysis of the collective mode spectra}}

\noindent
Despite the fact that both charge and spin
CMs couple to the QPs in the same hot-spot regions of the Brillouin zone,
symmetry arguments imply that their different characteristic wavevectors gives rise to
important cancellations in the memory function depending on the channel form factors.
The cancellation occurs when the relevant scattering wavevectors connect regions where the
form factors have the same sign (see Eq. (\ref{S1}) in the Appendix).
Conversely the largest contribution is obtained when the
critical wavevectors connect regions with opposite sign of the form factors.
Inspection of Fig. \ref{cmsr}
shows that the spin CM dominates the $B_{1g}$ spectra, while
the charge CM dominates the spectra in the $B_{2g}$ channel.
These ``selection rules'' are effective at low frequency and gradually fail
upon increasing the energy of the Raman scattering.
This failure is effectively taken into account by a gradual frequency-dependent switching on
(from  about 1000 cm$^{-1}$) of the coupling between the QPs and the ``forbidden''
CM in each channel (for details see Appendix ).
We do not attempt to fit the data above 4000 cm$^{-1}$ because multiple scattering processes
become important making the distinction between $M_c$ and $M_s$ meaningless.
At this point the spectra eventually become similar in shape [see Fig. \ref{fig1}(a)].

\begin{figure*}
\includegraphics[angle=0,scale=1.]{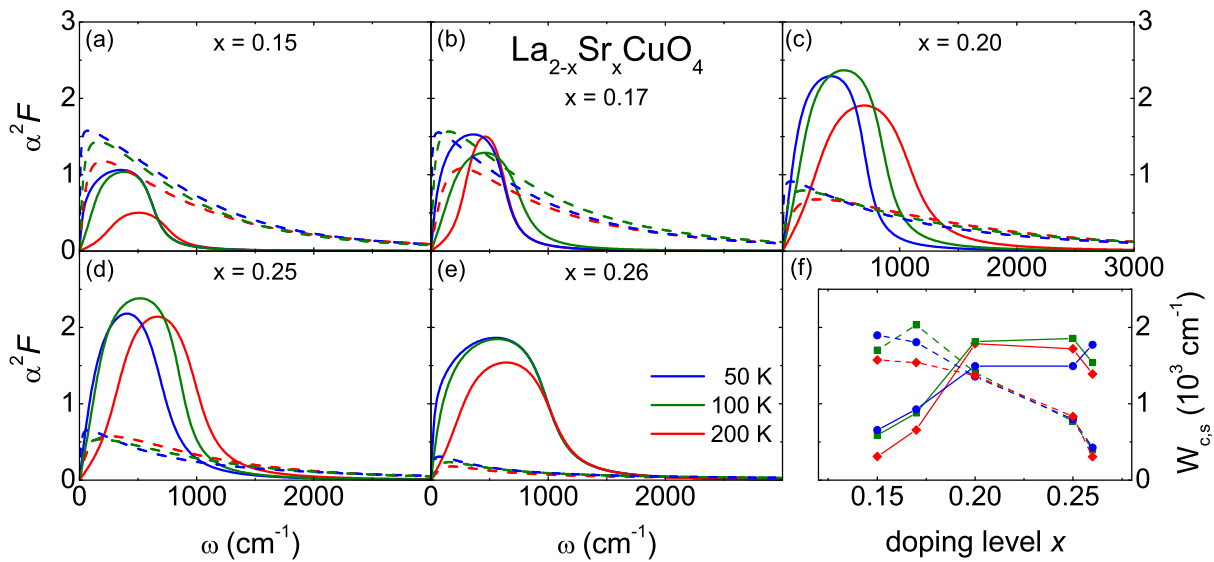}
\caption{(a)-(e) Glue functions due to the ``allowed'' modes in the fitted spectra of
Fig. 1. The solid lines are the glue functions due to charge CM, while the dashed
lines are the spin CM glue functions. Blue, green and red curves refer to low ($T\sim 50$ K),
intermediate ($T\sim 100$ K), and high ($T\sim 150-200$ K) respectively.
(a) $x=0.15$, $T=$42 K ,95 K, and 195 K; (b) $x=0.17$, $T=$54 K, 104 K, 254 K; (c)
$x=0.20$, $T=$56 K, 104 K, 201 K; (d) $x=0.25$, $T=$53 K, 104 K, 201 K;   (e) $x=0.26$ $T=$46 K,
102 K, 149 K; (f) Total weight of the glue functions mediated by charge
$W_c$ (solid lines) and spin  $W_s$ (dashed lines) as a function of doping $x$}
\label{fig4}
%
\end{figure*}

Adjusting the CM parameters and their coupling to the QPs we fit the experimental spectra (see
Fig. \ref{fig1}). Remarkably, the different shapes and
doping evolutions can be tracked. For instance, the
$B_{1g}$ spectrum evolves from the rounded shape (characteristic of a strong
scattering already at low energies) at $x=0.15$ to the narrow peak typical
of a weakly scattered ``Drude-like'' behaviour at $x=0.26$.
From these systematic fits we extract the corresponding evolution of the
glue functions (see Fig. \ref{fig4}) and of the related CMs [see Fig. \ref{fig2}(b-d)].
Quite naturally in the substantially doped regime we are considering, we find
that the spin spectra (dashed lines) have a marked diffusive character
(typical values of $\OO$ are in the range of $1000-3000$ cm$^{-1}$).
The overall weight $W_s$ of the glue functions due to spin exchange [see the
dashed lines in Fig. 4(f)]
is still rather large at $x=0.17$, but then decreases and becomes
very small in the most overdoped sample ($x=0.26$) in agreement with inelastic
neutron scattering data \cite{wakimoto1,wakimoto2}.
Conversely, the shape of the charge-mediated glue is more propagating-like
(roughly resembling the shape proper to a marginal-Fermi-liquid glue) because the typical $\OO\sim 200$
cm$^{-1}$ is much lower than in the spin case. Thus, the charge-mediated glue function
is centered at typical phonon frequencies of the cuprates. This agrees well with the idea that
charge-ordering in cuprates is phonon-driven \cite{CDG}, with a strong mixing between the electronic
and the lattice degrees of freedom \cite{BTGD}. Charge CMs have a marked phononic
character at generic $\qvec$ and acquire a strong electronic component at $\qvec \sim \qvec_c$,
where this composite excitation has an anomalous softening with weight displacement down to
low energies of order $m_c$. This behaviour is reflected in the temperature dependence of
the charge glue functions, which, while having their centre at about
$500$ cm$^{-1}$, also have a substantial spectral density at low energy when temperature is low.

The overall weight of the charge glue $W_c$ evolves in the opposite way with respect to
the spin weight. It is relatively small at $x=0.15$, increases reaching a maximum at
$x=0.20-0.25$ and then slightly decreases at $x=0.26$
[see Fig. 4(f)]. Of course this doping evolution should be reflected
in the doping evolution of the superconducting critical temperature. Owing to the
strongly retarded character of the interaction, a full Eliashberg analysis
of superconducting properties is required.

Both charge and spin CMs have strongly temperature dependent inverse square correlation lenghts
$\xi^{-2}=m/\Lambda$ (see Figs. 2(b),(c)) which only affect the low-frequency 
part of the glue functions,
but have a marked influence on the Raman spectra over the whole frequency range. 
Without this temperature
dependence (like, e.g., for standard phonon scattering), the spectra would be much more temperature
dependent, due to the bosonic occupation factors in Eq. (\ref{immemory}).

In general, we find that the observed temperature dependence 
of the charge inverse square correlation length
$\xi^{-2}_c$ is in qualitative agreement with an underlying critical behaviour
related to a competing charge-ordering instability. In particular, we notice that 
the linear temperature dependence
at $x=0.20$ extrapolates to zero  at  $T\to 0$ indicating that
this doping is quite close to the value of the quantum critical point usually reported to be
at $x=0.19$ \cite{CDGZP,reviewQCP1}).

We also find that the spin correlation length is of the same order as the correlation of the charge
ordering at low $T$, consistent with the idea that spin fluctuations can survive at high doping, far
from the antiferromagnetic region, if sustained and ``enslaved'' by charge-depleted fluctuating 
regions.
This connection is also apparent in Fig. \ref{fig2}(d), where a similar energy scale $\omega_1$ for
charge and spin is reported. This shows that the ``bulk'' of the spectrum
for the two collective degrees of freedom has essentially
the same energy threshold.

\vspace{0.5 truecm}
\noindent

\section{Other experiments}

\noindent
Coupling functions derived recently from the Raman spectra in one Bi2212 sample 
yield differences in the two symmetries,
with a prominent spin scattering in the $B_{1g}$ channel \cite{Muschler:2010b}. 
However, the identification of spin and charge modes and the assessment of their contribution
to the two channels at low frequency has been made possible by the procedure described here.
The present work may solve the long-standing puzzle in the 
cuprates regarding the relevance of the spin vs. phononic
scattering mechanism affecting, e.g., the electron dispersion and giving rise to the kinks
in ARPES spectra \cite{lanzara1,lanzara2,chubukov}:
 both spin and mixed charge-phonon modes contribute to the anomalies of the
electronic dispersion because both modes share the same hot-spot region and have a substantial weight.
It is also encouraging that recent ARPES \cite{hashimoto} and STM \cite{howald,hanaguri,kohsaka}
experiments support the role of charge-order scattering in electronic spectra.
Moreover, the relevance of charge-order excitations with a marked phononic character, can also
provide a rationale for various isotopic effects detected along the years in cuprates
\cite{jlee,hofer,rubiotemprano,takagi,andergassen}

\begin{figure}
\includegraphics[angle=0,scale=1.]{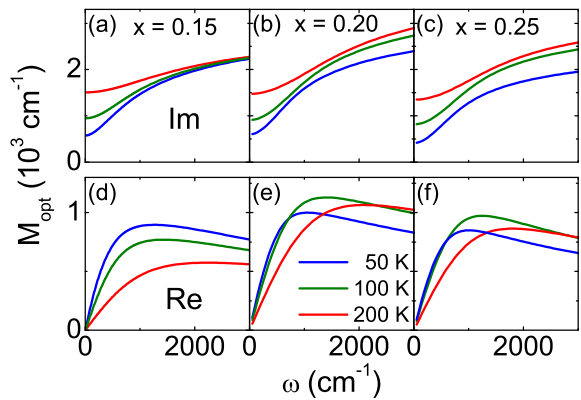}
\caption{Imaginary (a-c) and real (d-f) parts of the optical memory functions
obtained summing the charge and the spin glue functions reported in Fig. \ref{fig4}.
The correspondence between colors and temperatures is the same as in Fig. \ref{fig4}.
For details of the calculations see the Appendix }
\label{fig5}
%
\end{figure}

The simultaneous presence of two CMs dressing the QPs, was also recently found
in optical spectroscopy on Bi 2201 and Bi 2212 samples \cite{yang,vanheumen}. One could also 
 similarly interpret the $\alpha^2F$ recently obtained from ARPES spectra \cite{schachinger,bok}. These 
findings are in agreement with our analysis, because in optical
conductivity and in ARPES the channel-dependent form factors are absent and
both CMs contribute. As a result we simply need to add the spin and charge
glue-functions to reproduce the experiments. Although we did not attempt a precise fitting
and comparison with the experimental data, we explain why the optical scattering time
($Im\,M^{\rm opt}(\omega)$) above 1000 cm$^{-1}$ displays
a very weak (if not absent) temperature dependence in underdoped samples, which
gradually increases upon doping (see Fig. \ref{fig5}(a)-(c))\cite{timusk,puchkov,hwang}.
This behaviour simply arises from the temperature dependence of the CM masses (both charge
and spin): upon increasing $T$, the masses grow producing a less marked increase of the
scattering. This compensation is less effective upon increasing doping.

While the above spectroscopies indirectly probe the CM through their effects of
the QPs, other spectroscopies directly probing the charge or spin CMs provide pictures  consistent
with our findings.  
Concerning the spin modes, there is a
correspondence between our spin glue functions and the
spin spectral functions obtained from inelastic neutron scattering. Specifically
we found that above $x=0.20$ the spin excitation intensity rapidly decreases, by
about a factor one half between $x=0.25$ and $x=0.26$. This corresponds well to the
spin suppression observed in Refs. \onlinecite{wakimoto1,wakimoto2}.
Dynamical charge fluctuations are less accessible. However, we interpreted previous results of Raman
absorption in underdoped LSCO samples at low frequencies (up to a few hundreds of
wavelengths) \cite{tassini} in terms of direct excitations of two charge CMs \cite{suppa}. 
Remarkably the values
and the temperature dependencies of $m_c(T)$ obtained at such low frequencies and in
much less doped samples, and from different excitation processes, are in close agreement
with the behaviour of $m_c(T)$ found in the presently analyzed optimally and overdoped samples.

\vspace{0.5 truecm}
\noindent
\section{Summary and perspectives}

\noindent
In this paper we demonstrate that a retarded bosonic glue accounts well
for the rich variety of shapes, and of doping and temperature dependencies
of Raman spectra in LSCO.
We show that, fully exploiting the interplay between the sign
of the Raman form factors and the peaked momentum structure of the fluctuations,
the dominant absorption in the $B_{1g}$ and
$B_{2g}$ channels at low energy mostly arises from QPs
scattering either with spin or charge
CMs respectively. The different shapes of the Raman spectra  in $B_{1g}$ and $B_{2g}$ symmetry
is the result of the more diffusive or propagating character
of the corresponding dominant CMs.
The more propagating charge CM corresponds to a rather
flattish glue function thereby reproducing the phenomenology of marginal-Fermi-liquid
theory in momentum integrated responses. 

Our main result is a complete description of
doping and temperature evolution of the glue functions due to charge and spin CMs.
The relative importance of the two scattering
mechanisms switches from spin to charge by increasing doping.
While the relevance of dynamical spin scattering was already inferred both from theoretical approaches
\cite{hanke}  and phenomenological analyses \cite{dahm}, 
the simultaneous presence of mixed phonon-charge-density fluctuations 
and their increasing relevance with doping is assessed here for the first time. 
Moreover, the relevance of charge fluctuations also indicates that a nearly charge-ordered phase competes
with superconductivity in underdoped LSCO.
Our analysis opens the way to the wide field of investigations based on a specific
form of a retarded bosonic glue.

\noindent {\bf Acknowledgments}

\noindent
We acknowledge discussions with C. Castellani, T. Devereaux, and J. Lorenzana.
S.C., C. D.C., and M.G acknowledge financial support form the
 MIUR-PRIN07 prot. 2007FW3MJX\_003.
R.H. and B. M. acknowledge support from the DFG under grant-no. Ha2071/3 via Research Unit FOR538.
 M.L. and A.E. were support in the same network under grant number Er342/1.
\appendix

\section{Memory function}
\begin{figure}
\includegraphics[angle=90,scale=0.3]{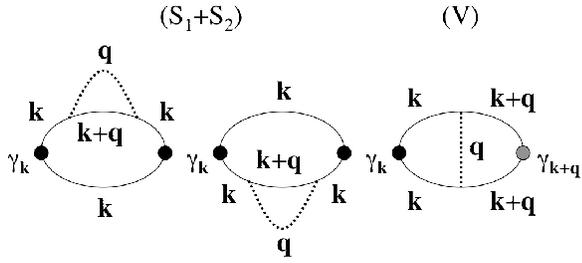}
\caption{Self-energy (S) and vertex (V) diagrams corrections to the Raman response
function with the exchange of a single
collective mode. The black dots represent the $\gamma_\kvec$ form factor, while the shaded
dot represents $\gamma_{\kvec+\qvec}$. The dotted lines represent collective
mode Green's functions}
\label{sifig1}
%
\end{figure}

We outline here the main steps of the calculation of the memory function
that appears in the Raman response equation (2). To simplify the notation we drop
the index $\mu$, since the formal expressions are the same for all symmetry channels.
Within a perturbative approach limited to the exchange of a single CM, the memory function reads
$$
M=-\omega\frac{\chi_{SV}}{\chi_0},
$$
where $\chi_0$ is the purely real low-frequency Raman response function in the
absence of any scattering process. 
and $\chi_{SV}$ is the contribution due to CM selfenergy (S) and vertex (V) corrections
(see Fig. \ref{sifig1}). Here we are neglecting impurity scattering as well as  
sources of scattering other than CMs.
The fermion loops entering the expression of
$\chi_{SV}(\Omega_M)$, where $\Omega_M$ is the external Matsubara frequency, are
\begin{eqnarray*}
S_1({\mathbf q},\omega_n,\Omega_M)&=&-2T\tilde g^2\sum_{\epsilon_\ell}\int\frac{d^2\mathbf k}{(2\pi)^2}
\gamma_{\mathbf k}^2
\mathcal{G}({\mathbf k},\epsilon_\ell-\Omega_M)\nonumber\\
&\times&\mathcal{G}({\mathbf k},\epsilon_\ell)\mathcal{G}({\mathbf k}+{\mathbf q},
\epsilon_\ell+\omega_n)\mathcal{G}({\mathbf k},\epsilon_\ell),\nonumber\\
S_2({\mathbf q},\omega_n,\Omega_M)&=&-2T\tilde g^2\sum_{\epsilon_\ell}\int\frac{d^2\mathbf k}{(2\pi)^2}
\gamma_{\mathbf k}^2
\mathcal{G}({\mathbf k},\epsilon_\ell+\Omega_M)\nonumber\\
&\times&\mathcal{G}({\mathbf k},\epsilon_\ell)\mathcal{G}({\mathbf k}+{\mathbf q},
\epsilon_\ell+\omega_n)\mathcal{G}({\mathbf k},\epsilon_\ell),\nonumber\\
V({\mathbf q},\omega_n,\Omega_M)&=&-2T\tilde g^2\sum_{\epsilon_\ell}\int\frac{d^2\mathbf k}{(2\pi)^2}
\gamma_{\mathbf k}
\gamma_{{\mathbf k}+{\mathbf q}}\mathcal{G}({\mathbf k},\epsilon_\ell)\nonumber\\
&\times&
\mathcal{G}({\mathbf k},\epsilon_\ell+\Omega_M)
\mathcal{G}({\mathbf k}+{\mathbf q},\epsilon_\ell+\omega_n+\Omega_M)\nonumber\\
&\times&
\mathcal{G}({\mathbf k}+{\mathbf q},\epsilon_\ell+\omega_n),\nonumber
\end{eqnarray*}
where
${\mathcal G}({\mathbf k},\epsilon_\ell)=(i\epsilon_\ell-\xi_{\mathbf k})^{-1}$
is the Matsubara Green's function of free fermions with band dispersion
$\xi_{\mathbf k}$.
To calculate $\chi_{SV}$, we make now two main approximations. First,
we notice that the CM propagator is peaked at a wavevector
$\mathbf{q}=\mathbf{q}_\lambda$, whereas the fermion loops are smooth functions
of the integrated wavevector $\mathbf{q}$. Thus, in the integral over
$\mathbf{q}$ that appears in the expression for $\chi_{SV}$, the fermion loops can be 
factorized and evaluated at $\mathbf{q}=\mathbf{q}_\lambda$. Once this factorization 
is introduced, the
integral over $\mathbf{k}$ in the fermion loops is dominated by QPs near the
hot spots (HS)at the Fermi surface associated 
with $\mathbf{q}_\lambda$. The two-dimensional integral over 
$\mathbf{k}$
is then suitably transformed into an integral over the QP dispersions at the two
hot spots connected by $\mathbf{q}_\lambda$, $\xi_{HS}$ and
$\xi^{\prime}_{HS}$ (this approximation corresponds to a linearization of
the band dispersion around the hot spots, see Ref. \cite{chubukov}). This 
change of variables
involves the Jacobian of the transformation $J$ evaluated at the hot spots,
giving rise to a dimensionless prefactor $\tilde{J}\equiv J\tilde{g}^2$.
The subsequent sum over the fermion Matsubara
frequency $\epsilon_\ell$ may be carried out explicitly.

In the forthcoming discussion, it is crucial to observe that the
selfenergy contribution contains two Raman form factors evaluated at the
same hot spot, $\gamma^{HS}$, whereas one of the two form factors appearing
in the vertex contribution, $\gamma^{HS'}$, is evaluated at a
different hot spot, connected to the former by $\mathbf{q}_\lambda$.

After evaluating the fermion loops, within the approximations outlined
above, we obtain the contribution to the Raman response
$$
\chi_{SV}(\Omega_M)=T\sum_{\omega_n}\Gamma_{SV}(\omega_n,\Omega_M)
\int\frac{d^2\mathbf{q}}{(2\pi)^2}
\mathcal{D}(\mathbf{q},\omega_n),
$$
where $\mathcal{D}(\mathbf{q},\omega_n)=-[m_\lambda+\nu_\lambda(
\mathbf{q}-\mathbf{q}_\lambda)^2+|\omega_n|+\omega_n^2/\bar\Omega_\lambda]^{-1}$
is the Matsubara CM Green's function and the leading contribution of
the fermion loops is fully included in the function
$\Gamma_{SV}(\omega_n,\Omega_M)\equiv\left[
S_1({\mathbf q},\omega_n,\Omega_M)+S_2({\mathbf q},\omega_n,\Omega_M)+V({\mathbf q},\omega_n,\Omega_M)
\right]_{{\mathbf q}=\mathbf{q}_\lambda}$. 

By explicit evaluation, following the procedure outlined above, we find
\beq
\Gamma_{SV}(\omega_n,\Omega_M)=4\pi\tilde{J}\gamma^{HS}\left(
\gamma^{HS'}-\gamma^{HS}\right)\frac{\left|\omega_n\right|-
\left|\Omega_M\right|}{\Omega_M^2}\nonumber \,\,\, \label{S1},
\eeq
for $-|\Omega_M|<\omega_n<|\Omega_M|$, and $\Gamma_{SV}(\omega_n,\Omega_M)=0$
elsewhere. This expression enforces the selection rule quoted in the text:
the main contribution to the fermion loops is non vanishing only if the
Raman form factor $\gamma_{\mathbf k}$ has a different sign on the two hot
spots connected by $\mathbf{q}_\lambda$. For the characteristic wavevectors
of spin and charge CMs, this occurs in the $B_{1g}$ and $B_{2g}$ symmetry
channel, respectively.

After a shift of the origin of momentum space to $\mathbf{q}_\lambda$, the
integral over $\mathbf{q}$ becomes straightforward. The sum over the
Matsubara frequencies $\omega_n$ is then transformed into an integral over
the real axis by standard techniques, exploiting the fact that the boson
frequencies $\omega_n$ are the simple poles of $\coth(z/2T)$.
After causal continuation of the
external Matsubara frequency $\Omega_M$ to the real frequency $\omega$, we
finally obtain the contribution to the Raman response function
due to the QP scattering with CMs,
\begin{eqnarray*}
\chi_{SV}(\omega)&=&-\frac{\tilde{J}\gamma^{HS}\left(
\gamma^{HS}-\gamma^{HS'}\right)}{2\pi\nu_\lambda\omega^2}
P\int_{-\infty}^{+\infty}dz\nonumber\\
&\times&\left[\frac{1}{2}\log\frac{\bar\Omega_\lambda^2z^2+
(\bar\Omega_\lambda\Lambda_\lambda-z^2)^2}{\bar\Omega_\lambda^2z^2+(\bar\Omega_
\lambda m_\lambda-z^2)^2}
\right.\nonumber\\
&+&\left. i\left(\arctan\frac{\bar\Omega_\lambda
\Lambda_\lambda-z^2}{\bar\Omega_\lambda z}
-\arctan\frac{\bar\Omega_\lambda m_\lambda-z^2}{\bar\Omega_\lambda z}\right)\right]
\nonumber\\
&\times&(\omega-z)\left(\coth\frac{z}{2T}-\coth\frac{z-\omega}{2T}\right),
\nonumber
\end{eqnarray*}
where the symbol $P$ indicates that the principal part of the integral
must be considered whenever a simple pole of the integrand is met on the
real axis. Although this is not immediately evident, by changing the integration
variable to $-z$ and summing one half of each of the two expressions
it may be easily shown that the imaginary part of the resulting memory function is an 
odd function of the external frequency $\omega$, as required by causality. This explicit
antisimmetrization with respect to $\omega$ yields the results for $Im M$
and $\alpha^2F$ quoted in the text, with the dimensionless effective
coupling
$$
g_\lambda=\frac{\tilde{J}\gamma^{HS}\left(\gamma^{HS}
-\gamma^{HS'}\right)}{2\pi \nu_\lambda \chi_0}.
$$
In the presence of multiple hot spots and/or multiple values of the
characteristic wavevectors $\qvec_\lambda$ (as implied, e.g., by the symmetry
of the lattice), the overall dimensionless coupling is intended to be a sum
over all the hot spots and $\qvec_\lambda$'s.

\section{Shape of the Raman response}

The connection between the overall shape of the Raman response and the
properties of the CMs is fully entailed in the form of the CM glue function
$\alpha^2F(\omega)$. In the forthcoming discussion it is important
to observe that the overall behavior of the Raman response is
qualitatively understood by inspecting the behavior of the frequency
dependent inverse scattering time $1/\tau(\omega)=Im M(\omega)$, since
the mass corrections associated to $Re M(\omega)$ do not introduce
significant features.

\begin{figure}
\includegraphics[width=6cm]{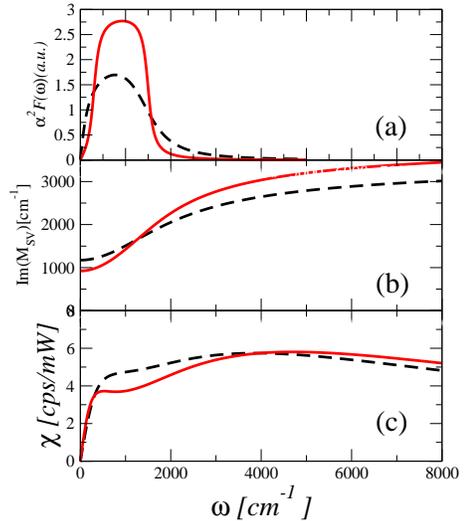}
\caption{(a) Glue functions for two CMs having the same dynamical range between
$\omega_1=330$ cm$^{-1}$ and $\omega_2=1500$ cm$^{-1}$, but different
diffusive-propagating crossover scale: $\OO=200$ cm$^{-1}$ for the more propagating
mode (solid red curve), $\OO=1000$ cm$^{-1}$ for the more diffusive
mode (dashed black curve). (b) Imaginary part of the memory function (scattering time)
corresponding to the two glue functions in (a). (c) Raman response obtained for the
two glue functions in (a) with $g^2=0.5$.}
\label{sifig2}
%
\end{figure}

A nearly constant $1/\tau$ produces a Drude-like peak at frequency
$\omega\approx 1/\tau$. A nearly linear $1/\tau$ produces instead a
flattish (i.e., marginal-Fermi-liquid-like) response. Now, we see that a
diffusive CM (large $\bar\Omega$) is characterized by a broadly peaked
glue function [see black dashed line in Fig. \ref{sifig2}(a)].
This, in turn, yields a $1/\tau$ which smoothly interpolates
between the low-frequency value ($1/\tau_0$) and the high-frequency value
($1/\tau_\infty>1/\tau_0$)
[see black dashed line in Fig. \ref{sifig2}(b)].
The Raman spectrum is correspondingly characterized
by a hump+hump overall shape which results from the overlap of two Drude
peaks roughly centered at $1/\tau_0$ and $1/\tau_\infty$
[see black dashed line in Fig. \ref{sifig2}(c)]. This corresponds
to the $B_{1g}$ spectra experimentally observed in LSCO.
If the CM has a marked propagating character (small $\bar\Omega$), the
corresponding glue function has instead the shape of a rounded box function
[see red solid line in Fig. \ref{sifig2}(a)]. This, in
turn, implies a broad linear regime in $1/\tau$ and a wider separation between
$1/\tau_0$ and $1/\tau_\infty$. The corresponding Raman spectra has the
step+hump form [see red solid line in Fig. \ref{sifig2}(c)],
similar to the $B_{2g}$ spectra experimentally observed in  LSCO.

\section{Relaxing the selection rules}

The channel-dependent selection rule for the Raman spectra
was obtained under two main assumptions: (1) the CMs are peaked
at finite wavevectors $\mathbf{q}_\lambda$ and (2) the factorized
fermion loops are dominated by QPs at the hot spots. The second
assumption is asymptotically valid at low frequency, but is gradually
violated at higher frequency. This means that at sufficiently large
$\omega$ both CMs contribute to Raman spectra, regardless of the
symmetry. This is expected on theoretical grounds and indeed
observed experimentally: the $B_{1g}$ and $B_{2g}$ spectra of LSCO
have similar shapes at high frequency, indicating that channel-dependent
selection rules are not at work there.

However, the analytical evaluation of the fermion loops beyond the
hot-spot contribution is not viable and the numerical evaluation is
both demanding and not transparent. Thus, in our calculation we adopted
an alternative phenomenological description, writing the memory function
as $M(\omega)=M_a(\omega)+s(\omega)M_f(\omega)$, where $M_a(\omega)$ is
the memory function associated with the CM allowed by the selection rule
and $M_f(\omega)$ is the memory function associated with the CM forbidden
by the selection rule. The switch function
$$
s(\omega)=\frac{1}{2}\left[1+\tanh\left(\frac{\omega-\omega_f}{\beta_f}\right)
\right]
$$
gradually includes the contribution of the forbidden CM at a frequency
$\omega\approx\omega_f$. The rapidity of the inclusion is controlled by the
parameter $\beta_f$. Typical values in our fits are $\omega_f\sim 10^3$ cm$^{-1}$
and $\beta_f\sim 5\times 10^2$ cm$^{-1}$, roughly common to
both charge and spin modes.
We point out that our phenomenological procedure is not fully causal
(i.e., $Re M$ and $Im M$ are no longer related by
Kramers-Kronig transformations). However, since a constant $s(\omega)$
would restore causality, a sufficiently gradual inclusion of the
forbidden CM does not introduce significant drawbacks. Below the frequency $\omega_f$, the
spectra in the $B_{1g}$ and $B_{2g}$ symmetries are markedly different, whereas above
$\omega_f$ they gradually become similar in shape.


\end{document}